\documentclass[aps,prb,amsmath,amssymb,showpacs]{revtex4}

\usepackage[dvips]{graphicx}

\begin{document}
  \title{Step bunching process induced by the flow of steps at the sublimated crystal surface.}
\author{Magdalena A. Za{\l}uska--Kotur }
\email{zalum@ifpan.edu.pl}
 \affiliation{Institute of Physics, Polish Academy of Sciences,
Al. Lotnik{\'o}w 32/46, 02-668 Warsaw, Poland and Faculty of Mathematics and Natural Sciences,
Card. Stefan Wyszynski University, ul Dewajtis 5, 01-815 Warsaw, Poland}
 \author{Filip Krzy{\.z}ewski} 
\email{fkrzy@ifpan.edu.pl}, 
 \affiliation{Institute of Physics, Polish Academy of Sciences,
Al. Lotnik{\'o}w 32/46, 02-668 Warsaw, Poland} 

\begin{abstract}
Stepped GaN(0001) surface is studied by the kinetic Monte Carlo method and compared with the model based on Burton-Carbera-Frank equations.  Successive stages of surface pattern evolution during high temperature sublimation process are discussed. At low sublimation rates clear, well defined step bunches form. The process happens in the absence or for very low Schwoebel barriers at the ideal surface. Bunches of several steps  are well separated, move slowly and are rather stiff. Character of the process changes for more rapid sublimation process where double step formations become dominant and together with meanders and local bunches assemble into the less ordered  surface pattern. Solution of the analytic equations written for one dimensional system confirms that step bunching is induced by the particle advection caused by step-flow anisotropy. This anisotropy becomes important when due to the low Schwoebel barrier both sides of step are symmetric. Simulations show that in the opposite limit of very high Schwoebel barrier steps fracture and rough surface builds up. 
     \end{abstract}
\maketitle

\section{Introduction}
\label{sec:A}

Temperature sublimation from the  surface is the process in which crystal decays layer by layer. Such procedure is a part of crystal surface preparation to the epitaxial growth  \cite{weyher2,weyher3,nowak1,pearton,sato}, core of the contemporary nanotechnology procedures\cite{nakamura1, skierbiszewski1}.
Nanodevice performance depends on the quality of the synthesized layers, both in terms of crystallographic structure and of their chemical composition. In terms of the growth control, the relative thickness of the deposited layers is relatively small, therefore the results of the growth depend primarily on the structural and chemical state of the surface prior to epitaxy \cite{weyher2,weyher3,nowak1}. Therefore, preceding
epitaxial growth, the substrates are annealed to clean the surface\cite{weyher2,weyher3}.
Thus we would like to have full control of the surface pattern in this process. Surface patterns of various shapes are studied analytically and by numerical simulations both of the crystal growth and crystal sublimation processes. Depending on the assumed parameters step meandering, bunching or  surface roughening is observed \cite{misbah}. Results of many works devoted to analysis of  both process show variety of possible mechanisms and scenario of surface pattern evolution \cite{misbah,sato1,ranguelov,kato,kardar,bales,bena,uwaha,pimpinelli2,fok,Yagi}. 
We show that on annealing of quite simple (0001) vicinal face of the GaN crystal  steps at the surface  have a tendency of bunching rather than meandering. Step bunches are very often observed in experiments during Si, GaN or SiC growth \cite{Yagi,Xie,leroy,monch,Muller,ohtani} in most cases built by the  electromigrated particles  flow \cite{Yagi,Xie,leroy}. Numerically is was studied in the systems   with Ehrlich-Schwoebel (E-S) barrier \cite{sato3} in  the chemically etched systems \cite{pimpinelli,garcia,garcia2}, with some defects assumed \cite{kandel}, with electromigration flow of particles \cite{liu,stoyanov} or for curved steps \cite{israeli}. In this work we study clean surface with no defects included. Particle diffusion,  sublimation or step attachment  probabilities depend only on  the local bonding energy values. Surface  evolution is an effect of individual particle jumps and as a result steps join together  building bunches at the surface. It was shown that bunching can be caused by  Schwoebel barriers \cite{saito}, diffusion at step edge \cite{politi}, impurities \cite{kandel} or  driven particle flow \cite{misbah,sato1,ranguelov,dufay}. In most analytical  models step movement factor is neglected \cite{misbah,ranguelov,kato,kardar,bales,bena,uwaha}. However as it was discussed lately particle advection is in fact comparable with electromigration  or Schwoebel barrier asymmetry \cite{ ranguelov,dufay,krug,ivanov}. Results of our Monte Carlo (MC) simulations show that the step bunching  happens at the ideal stepped surface  with  Schwoebel barrier assumed  to be zero and no external  particle driving present.  To analyze the situation more closely we have written Burton-Carbera-Frank (BCF) equations for one-dimensional system without Schwoebel barriers and solved them numerically. The solution  in some range of parameters looks similar to this what we observe on simulating our 2D lattice gas model.
Next, terms accounting for the step flow in the equations were set to be zero. As a result steps stay well separated during system evolution and we observe  the tendency to  even up their relative distances. 
It can be understand that at sublimated surface  step bunching is caused by the mean particle flux, which flows downward across slowly moving steps. When we neglect step movement in the model steps do not bunch at all. 
Low Schwoebel barrier gives the possibility of  exchanging the particles between steps. Particle which detach from one step can  attach another if only their lifetime  at the surface is long enough. Results of our simulations show that when such exchange is blocked by too high barrier at the step the system stops evolving by step flow and rough surface builds up. In such case analytic one dimensional models based on BCF equations are not proper description of the step dynamics at the surface. In next Section  MC model of GaN(0001)is shortly presented  and results of simulations are discussed. In  Section III we show step evolution  described by BCF equations and compare it with the behavior of simulated system.

\section{Kinetic Monte Carlo simulations of step bunching process during crystal sublimation. }
\label{sec:B2}

Below we show step bunching process realized in the  MC simulated process within  the lattice gas model \cite{[33],[34],[35]}. We provide kinetic MC simulations for a system of GaN(0001) surface symmetry. At misoriented (0001) surface of GaN crystal every second layer is different. We study the system in the condition of N supersaturation. In such situation we assume that only  Ga atoms control all dynamic processes at the surface.
Two topmost layers of crystal particles are modeled. Particles detach from the other particles and start free diffusion over the surface.  From time to time particles are wiped out from the  crystal, realizing sublimation process.  Diffusion over the surface  happens with   probability 
\begin{equation}
\label{p_d}
D=D_0 e^{-\beta \Delta E},
\end{equation}
where $D_0=1$ is diffusion timescale and    
\begin{equation}
\label{eq:E}
\Delta E=\left\{\begin{array}{ll}
0,\quad \textrm{if $E_i(J)<E_f(J)$;} \\
E_i(J)-E_f(J),\quad \textrm{otherwise.} \\
\end{array} \right.
\end{equation}
is given by the initial $E_i(J)$ and the final $E_f(J)$ bonding energy of the jumping  atom.  $\beta=1/k_BT$ is temperature factor. Each jump of free particle along the surface is realized with the same probability factor \ref{p_d}. Jump rate strongly depends on the 
Ga - Ga interaction strength, which is  determined by the position of bonding atoms N. Interaction energy $E$ is larger, when  four Ga atoms form  closed tetrahedron around N atom in the middle. Each  particle belongs to  four different tetrahedrons, what leads to the particle energy
\begin{equation}
 \label{en_czastki}
E(J)=J\sum_{i=1}^{4}[\frac{1}{3}r \eta_i+(1-r)\delta_{\eta_i,3}].
\end{equation}
with a number of occupied neighboring sites within tetrahedron given by $\eta$.  
Parameter $J$ scales the energy of bonds, the sum in (\ref{en_czastki}) runs over four surrounding tetrahedrons, and $r$ describes the relative  strength of the four-body and the two-body interactions in the system. 

Particle diffuses over the surface until it escapes out of it. Desorption probability is given by
\begin{equation}
\label{D}
p_d=a_d e^{-\beta \nu}
\end{equation}
with timescale $a_d=1$ and the desorption potential $\nu$. The desorption rate is controlled via temperature $\beta$ and potential $\nu$. 

Surface is heated, and kept at the high temperature. During  the sublimation  process step patterns transform from one form to the other depending on the simulation time. We study evolution of this process. The initial configuration of the system consists with regular pattern of equally spaced steps  oriented in the $[1\bar210]$ direction. Due to the surface symmetry energy  structure of every second step is mirror image of the neighboring one. As a result mean velocity is the same for every step.
Heights of the neighboring terraces differ by one Ga atomic layer. Ga atoms detach from the step, diffuse along the terrace and desorb from the surface. Crystal height decays layer by layer. Periodic boundary conditions are applied in the lateral direction and helical in the direction in which the crystal grows. 
\begin{figure}
\includegraphics[width=5cm]{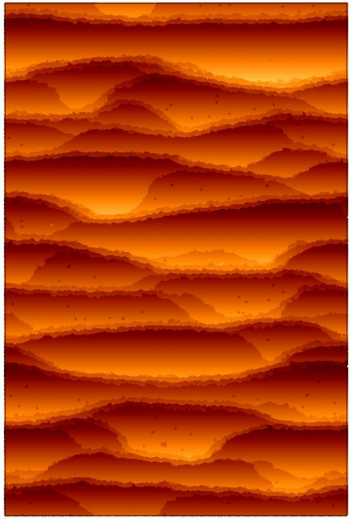}
\caption{(color online) 
Surface of simulated system with steps along  $[11\bar{2}0]$ direction . Size of the simulated system was $400a \times 600a$, $k_BT=0.22J$ and initial terrace width $d=10a$, $r=0.4$ , $B=0$ and $\nu=1.56J$. Parameter $s^2=0.09$ The surface after $7.5^.10^6 MC$ step evolution is shown, $25$ layers evaporated. } 	
\label{slow}
\end{figure}

\begin{figure}
\includegraphics[width=5cm,angle=0]{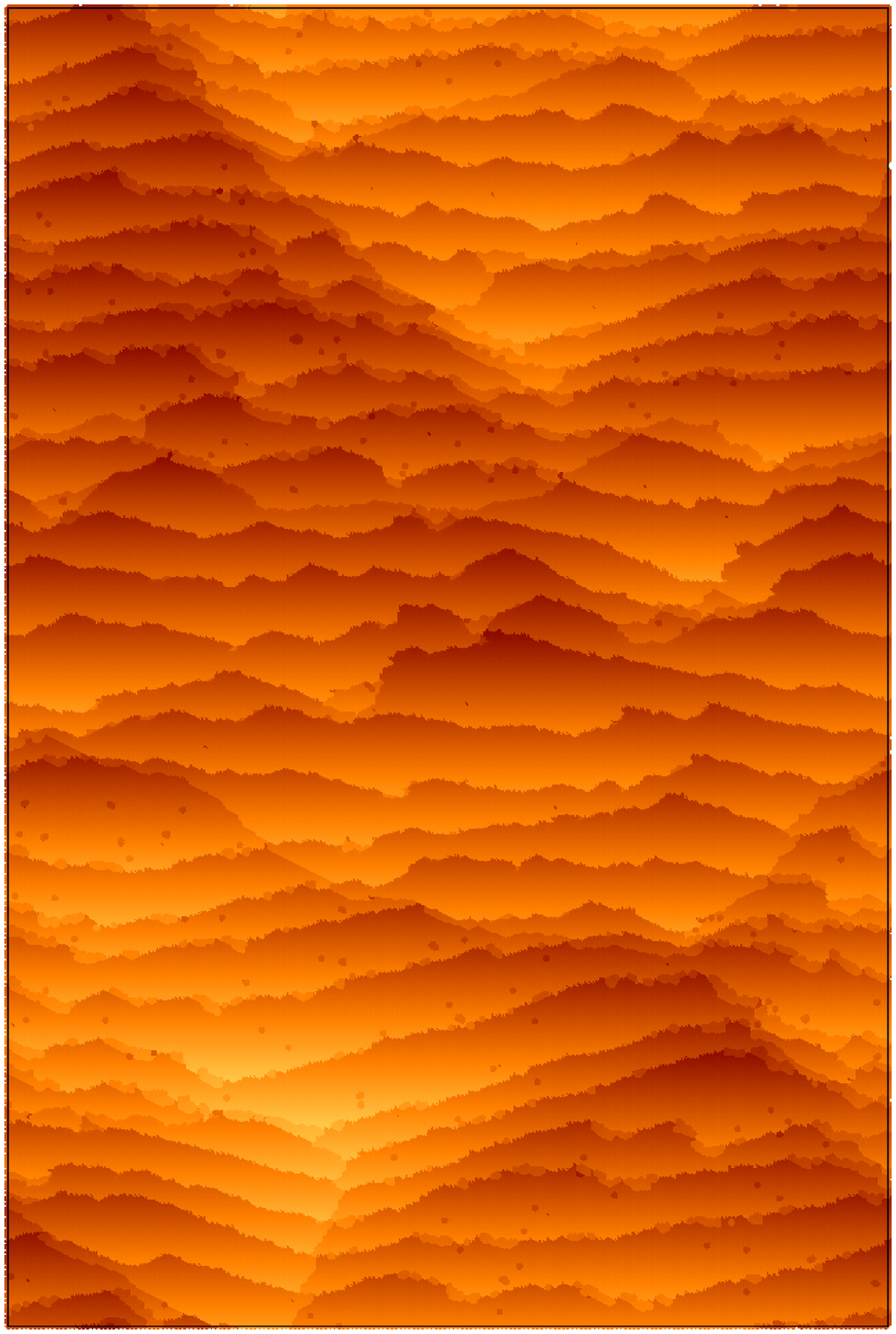}
\caption{\label{fast} (color online) Surface of simulated system with steps along  $[11\bar{2}0]$ direction . Size of the simulated system was $400a \times 600a$, $k_BT=0.22J$ and initial terrace width $d=10a$, $r=0.4$ , $B=0$ and $\nu=0.89J$. Parameter $s^2=1.9$. The surface after $3^.10^6 MC$ step evolution is shown, $120$ layers evaporated.}
\end{figure}

\begin{figure}
\includegraphics[width=7cm,angle=-90]{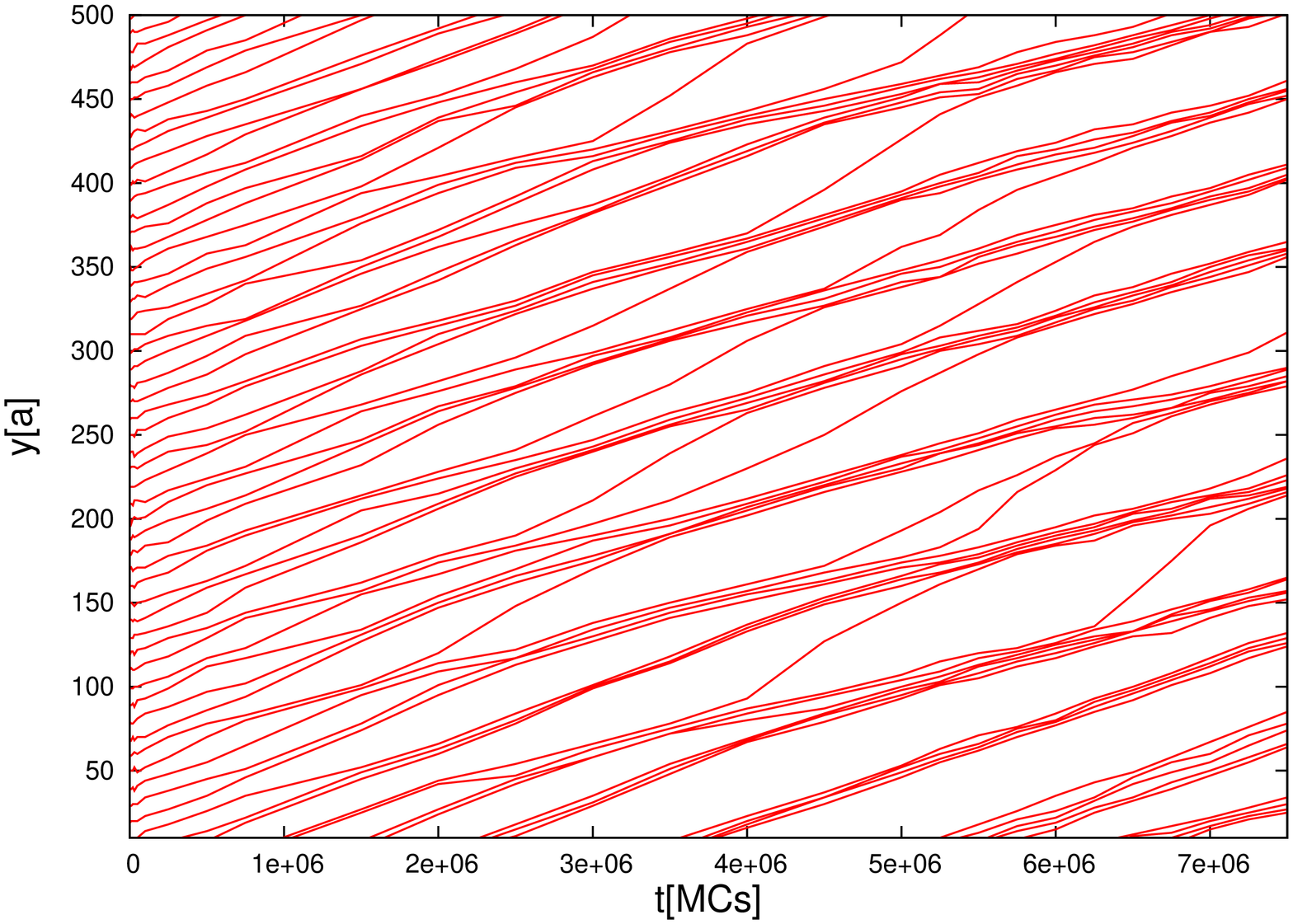}
\caption{\label{curly} (color online) Position of consecutive steps along $y$ axis as a function of time. At each time moment following step positions are measured at coordinate $x=200a$ for system from Fig.(\ref{slow})}
\end{figure}

\begin{figure}
\includegraphics[width=7cm,angle=-90]{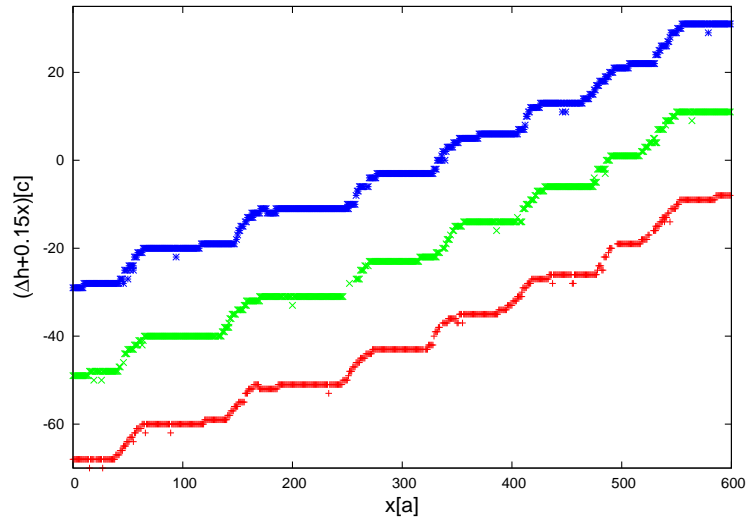}
\caption{(color online)Three different cuts of the step pattern (\ref{slow}) } 
\end{figure}

\begin{figure}
\includegraphics[width=7cm,angle=-90]{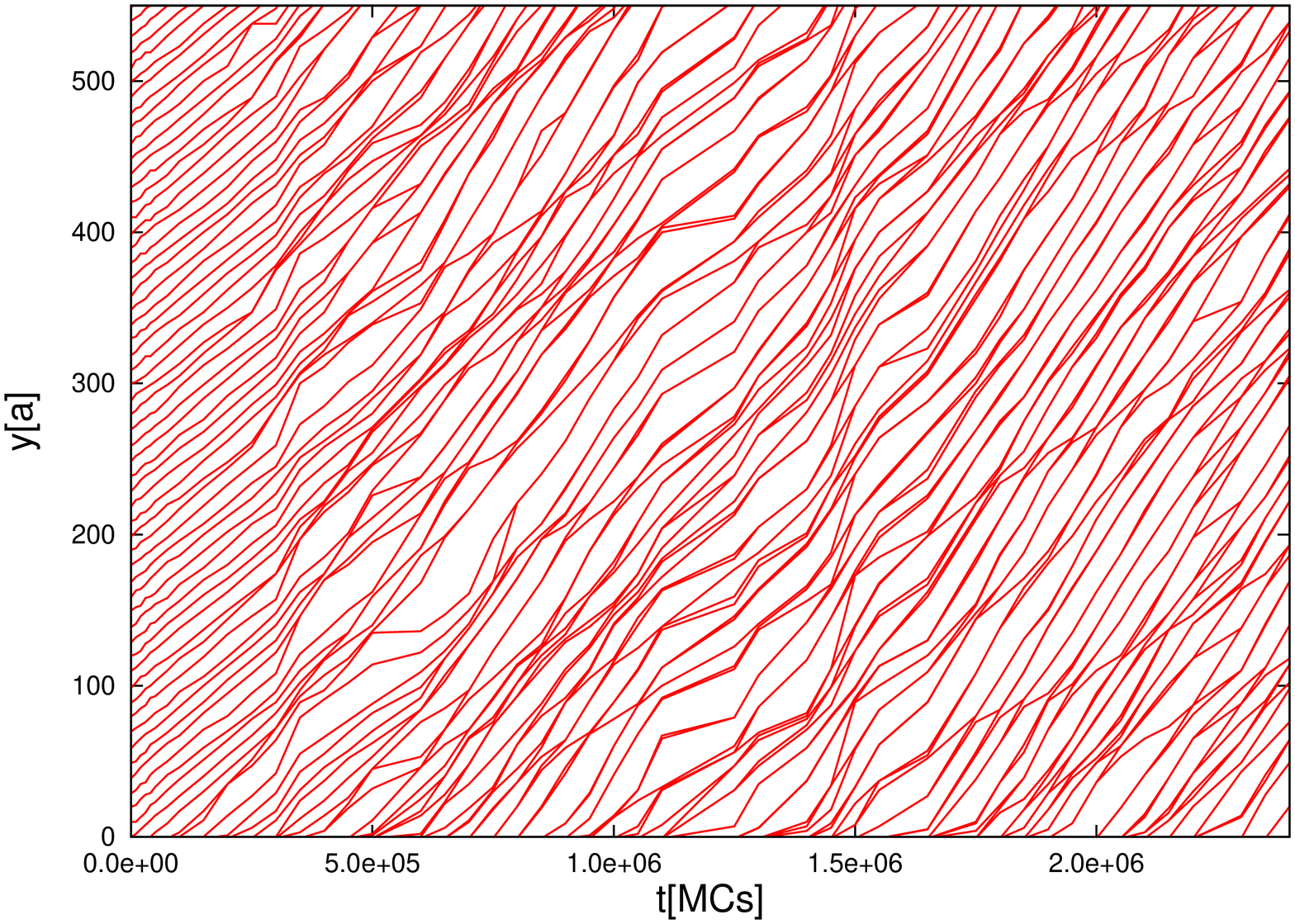}
\caption{(color online) Position of consecutive steps along $y$ axis as a function of time. At each time moment step positions are measured at coordinate $x=200a$ for system from Fig.(\ref{fast})
} 
\end{figure}
\begin{figure}
\includegraphics[width=5cm]{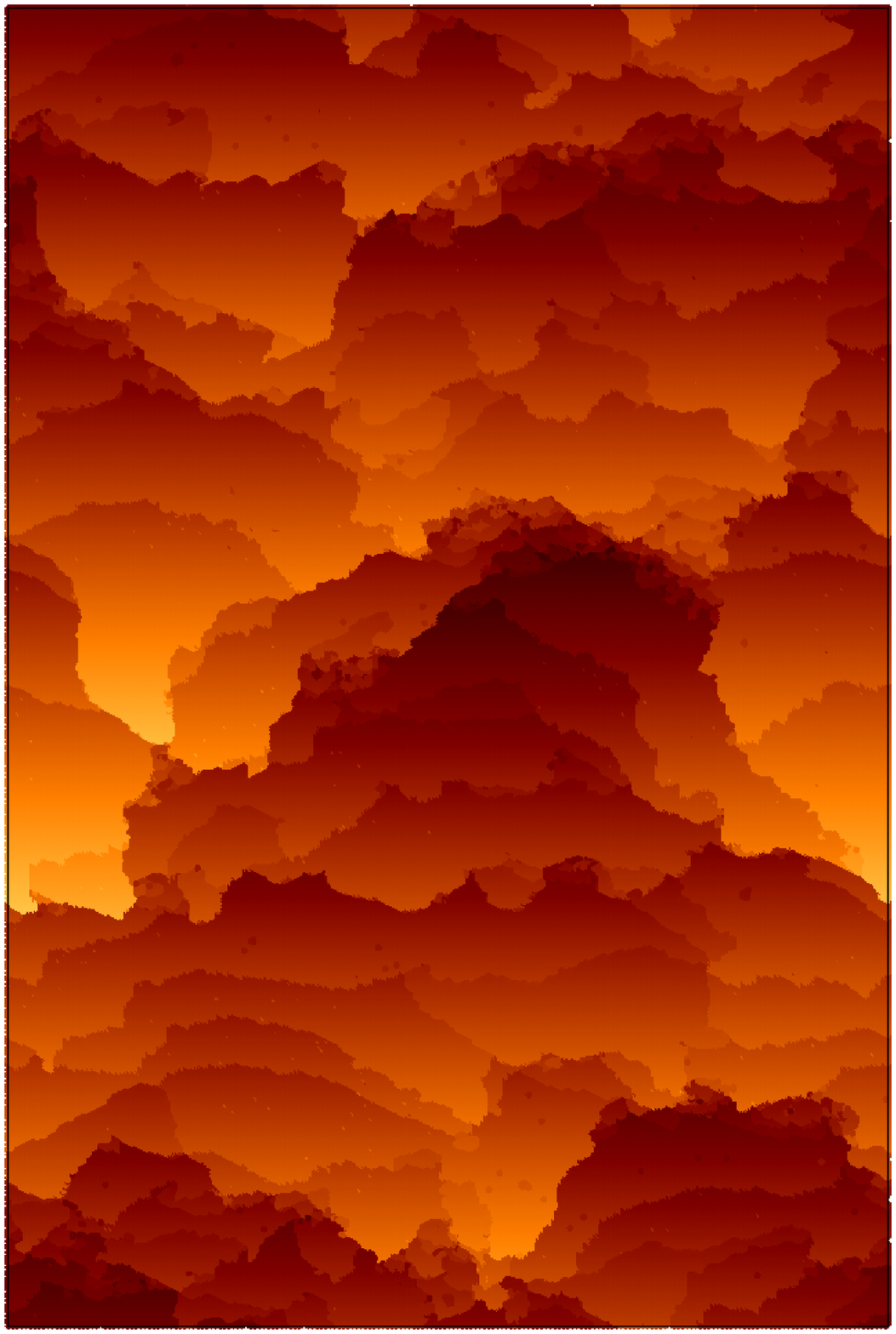}
\caption{(color online) 
Surface of simulated system with steps along  $[11\bar{2}0]$ direction. Size of the simulated system was $400a \times 600a$, $k_BT=0.22J$ and initial terrace width $d=10a$, $r=0.4$ , $B=1J$ and $\nu=1.56J$. The surface after $2.5^.10^5 MC$ step evolution is shown.
} 	
\label{schwoebel}
\end{figure}

During sublimation process steps move, bend and stick together. Successive step patterns transform smoothly each into another. Type of the pattern which finally emerge as a stable arrangement depends on the ratio between  mean time of the particle diffusion over the terrace of the width $L_0$ and the time it needs  to evaporate from the surface. Thus the important factor, which decides about the character of the step dynamics is 
\begin{equation}
s^2=\frac{p_d{L_0}^2}{D_0}={L_0}^2 e^{-\beta \nu}
\label{s2}
\end{equation}
on using formulas (\ref{p_d}) and (\ref{D}).  As long, as this ratio is less than one we have situation when part of particles evaporate from the terrace and part of them reach the next step and can attach it. In such a way steps communicate each with another, and global, ordered pattern can be build. When parameter $s^2$ is higher than one, it means that most of particles that wander over the surface evaporate, before they reach the  next step. Steps do not exchange many particles thus communication is weaker. Images in Figs \ref{slow} and \ref{fast} illustrate the effect of fast and slow sublimation process.
Step evolution for low value of $s^2=0.09$ is shown in Fig \ref{slow}. Regular, bunched structure is build. For higher but still close to one $s^2$ values  double step structures are formed and the resulting structure looks differently. In Fig \ref{fast}  pattern for $s^2=1.9$ is shown.   

When particles evaporate slowly from the surface part of them have a chance to attach to the next step. The number of attached  particles depends on the terrace width. The wider terrace is the less particles reaches next step. When they attach the step down of the  terrace, its width increases. In the case of upper step the situation is the opposite  and the terrace width decreases. When the number of  particles detaching and attaching at the step from the both sides is the same width of terrace stays stable. For zero Schwoebel barrier  at the step  both directions of jumps for each individual particle are  equally probable but the consequence  of particle detachment is that step moves backwards. Due to the step flow effective mean particle flux goes downward and breaks symmetry of the process. Particle flow down makes steps to stick together and group in bunches of several lines as seen in Fig \ref{slow}. At the wide terraces between them single steps can be found, bending on their way from one bench to the next. In order to look more closely at time evolution of this process we noted down positions of every  step at one arbitrary chosen cut of the surface plane in the direction vertical to steps and plot them as a function of simulation time. Resulting step position development  for the slowly sublimated system is plotted   in Fig 3. It can be seen that starting from regularly located  positions after some time steps stick together building pairs or triplets. These formations then join into larger groups of step bunches.  Already formed bunches exchange single steps, which move much faster as it is clearly seen in Fig 3. Comparing both figures: Fig \ref{slow} and  Fig 3 we see that single steps meander, hence their movement is a combination of translation and meandering. 
When step velocity grows with the distance  to the  neighboring steps  the contour of surface profile  has  characteristic convex form. Such convex shapes are apparent in Fig 4, where three different cuts of surface from Fig. \ref{slow} are shown. 

At higher rates of sublimation steps have tendency to stick in pairs rather than form larger step bunches. Most of the lines we see in Fig. 5 are double steps, both at the same location. It can be seen that they are created by two evolution lines sticking together. Some of double steps separate into two steps after some time. It can be seen in  Fig. 5 how  single lines detach from group of several steps down which means that they can be  slower than double steps. Overall system evolution  in this case is more chaotic than this visible in Fig. 3. Group of steps are smaller, double step formations are dominant and  no regular behavior can be seen. As a result such evolution leads to the surface structure Fig 2, which has different character than the one in Fig 1. The evaporation is so fast that diffusion along steps is not enough efficient to clear off all fluctuations, and as an effect steps shift their orientation forming meanders.  
The character of surface step dynamics changes with growing sublimation rate from the step bunching process seen in Fig. 3  to the less ordered one as in Fig 5.

Two cases discussed above are examples of the dynamics of the stepped surface without any Schwoebel barriers\cite{schwoebel} at the steps. Step bunching phenomenon is caused by the fact that step move, and there is net particle flux flowing downward. Interestingly when we add Schwoebel barrier $B$, changing diffusion across steps $D_B=D \exp(-\beta B)$, as long as $B$ is much smaller than the particle-step interaction, up to $0.6 J$, the character of the step bunching process does not change much.  When we come to higher Schwoebel barrier coming to one of order $J$ steps start to bend even if sublimation rate is low. In Fig. \ref{schwoebel} we see the surface of the system with Schwoebel barrier $B=J$ and annealed at the same condition as the system at Fig. \ref{slow}. It is even more rough than the one with fast sublimation process (Fig. \ref{fast}).   The step-step communication appears to be important for annealed two-dimensional system, similarly as in the case of grown crystal \cite{[34],[35]}.

\begin{figure}
\includegraphics[width=7cm,angle=-90]{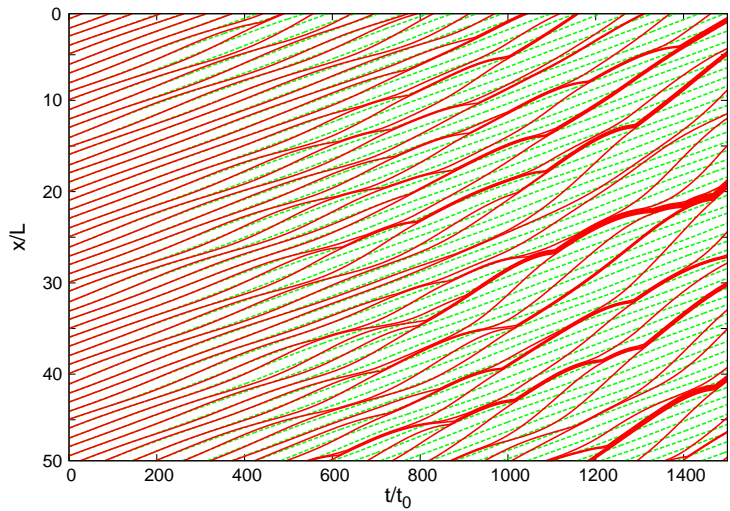}
\caption{(color online)Positions of 50 steps evolved according to eqs. (\ref{vn}) calculated for $s^2=0.09$, $\rho=5$, $C_0=0.4$, $a=10^{-9}$ are plotted in red line. Step bunches are formed. Single steps move from one bunch to the next one. Dashed (green) lines present results for the same system, when velocity term was neglected $v_0=0$.} 
\end{figure}

\begin{figure}
\includegraphics[width=7cm,angle=-90]{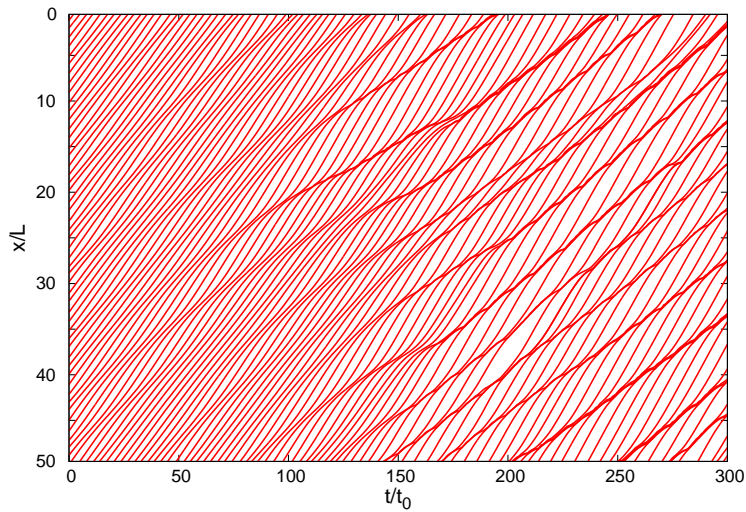}
\caption{ (color online) Positions of 50 steps evolved according to eqs (\ref{vn}) calculated for $s^2=1.9$, $\rho=5$, $C_0=0.4$, $a=10^{-9}$. Distances between steps oscillate and double step structures are build up at later stages of the evolution. } 
\end{figure}

\section{One-dimensional analytic model}

In most calculations on basis of Burton-Cabrera-Frank (BCF)\cite{misbah,ranguelov,kato,kardar,bales,bena,uwaha,BCF} model terms which contain step flow velocity are  neglected. Usually it does not change the results much especially when systems already have chosen one direction of preferred particle attachment or detachment flow either by step edge Schwoebel barrier, or by external flux. As it has been discussed  in Refs \onlinecite{dufay,popkov} particle advection due to step flow adds up to both these terms and moreover its strength in typical systems is comparable with them.  It affects the shape of  step stability phase diagram as shown in Ref. \onlinecite{dufay}. Our simulation results presented in previous Section  imply that in the absence of the Schwoebel step barrier or induced particle flux it is the step flow which leads to the step bunching. It should be  important when is the only term breaking left-right symmetry. To  study the connection between step flow and step bunching in more systematic way we analyzed one dimensional  BCF model  where the only anisotropy source is step movement. There is no step edge barrier and no external flux present. We check if  the step bunching phenomenon can be seen on the base of BCF equations.
We start with the basic expression 
\begin{equation}
\label{cn}
D_s \partial_{XX} C_n + V_0 \partial_X {C_n} -  {\frac {C_n} {\tau}}=0
\end{equation}
written for the adatom concentration $C_n$ at the $n$-th step in the frame moving with the mean step velocity $V_0$.
$D_s$ is the diffusion constant, $\tau$ is the mean particle lifetime at the surface  and we assume that steps descend in the positive direction of X. With such system orientation mean step velocity $V_0$ caused by the crystal sublimation is negative.  We recalculate coefficients of equation (\ref{cn}) by dividing it by $D_s$ and on using normalized  position $x=X/L_0$ and $l_n=L_n/L_0$, where $L_0$ is mean terrace width.  We use parameters $v_0=(V_0 L_0)/D_s$ as a relative mean step velocity and $s^2={L_0}^2/(\tau D_0)$ as a relative desorption rate. This last definition is exactly the same as  (\ref{s2}) on assuming $\tau={p_d}^{-1}$. Conditions at the step boundaries written in new quantities are given by 
\begin{eqnarray}
\label{boundary}
( \partial_x + v_0) C_n(0) = \rho (C_n(0)-{C^{eq}}_n) \nonumber \\
( \partial_x + v_0) C_n(l_n) = -\rho (C_n(l_n)-{C^{eq}}_{n+1})
\end{eqnarray}
with $\rho=(\nu L_0)/D_s$, where $\nu$ is step kinetic coefficient. Kinetic coefficient is assumed to be the same for both step sides, and ${C^{eq}}_n$ is the adatom equilibrium concentration. Step transparency is neglected. We also add step-step repulsive interaction, which modifies the equilibrium density
\begin{equation}
\label{ss}
{C^{eq}}_n = C_0+a(1/{l_n}^3-1/{l_{n-1}}^3)
\end{equation}
where $C_0$ is adatom equilibrium concentration at isolated step and $ a=A/{L_0}^3 $ measures step-step interaction coefficient of strength given by the constant $A$ \cite{Muller}. We use  $a= 1^. 10^{-9} $ which estimates entropic interaction between steps \cite{saito}  present also  in the lattice gas  simulation. We checked however that the final results do not depend much on the interaction given by (\ref{ss}). 

The solution of (\ref{cn}) can be written as 
\begin{equation}
\label{solution}
C_n=B_1 e^{r_1 x} +B_2 e^{r_2 x}
\end{equation}
with characteristic length coefficients
\begin{equation}
r_{1,2}=-\frac{v_0}{2} \pm \frac{1}{2}\sqrt{{v_0}^2+4s^2}= \frac{1}{2} (-v_0\pm \beta )
\end{equation}
After applying boundary conditions (\ref{boundary}) we can write solution for the step velocity
\begin{equation}
v_n= \rho(C_n(0)+C_{n-1}(l_{n-1})-2{C^{eq}}_n)
\end{equation}
which express by
\begin{eqnarray}
\label{vn}
\frac{v_n}{ \rho} = - {C^{eq}}_n \frac{(s^2+\frac{1}{2} v_0 \rho )\tanh(\frac{1}{2}\beta l_n)+\frac{1}{2}\beta\rho}{(s^2+\rho^2)\tanh(\frac{1}{2}\beta l_n)+\beta \rho}   
- {C^{eq}}_n \frac{(s^2-\frac{1}{2}v_0 \rho )\tanh(\frac{1}{2}\beta l_{n-1})+\frac{1}{2}\beta\rho} {(s^2+\rho^2)\tanh(\frac{1}{2}\beta l_{n-1})+\beta \rho}  \nonumber \\
+ {C^{eq}}_{n+1} \frac{ \frac{1}{2} \beta \rho e^{ \frac {1}{2} v_0 l_n } \cosh^{-1}(\frac{1}{2}\beta l_{n})} {(s^2+\rho^2)\tanh(\frac{1}{2}\beta l_{n})+\beta \rho} 
+ {C^{eq}}_{n-1} \frac{ \frac{1}{2} \beta \rho e^{-\frac{1}{2} v_0 l_{n-1} } \cosh^{-1}(\frac{1}{2}\beta l_{n-1})} {(s^2+\rho^2)\tanh(\frac{1}{2}\beta l_{n-1})+\beta \rho} 
\end{eqnarray}

Set of equations above is written for individual step velocities $v_n$.  Mean velocity value $v_0$ enters all equations. As long as we have stable flow of equidistant steps  with mean velocity $v_0$ Eq. (\ref{solution}) is  stationary solution of (\ref{cn}). When step velocities $v_n$ are different and vary with time the situation can change. Below we assume that these changes are not large, and study behavior of many steps in the approximation of the step density (\ref{solution}). Of course set of equally distanced steps is a stationary solution of equations (\ref{cn}). However, as it is well known for some parameters this solution is not stable  \cite{misbah,uwaha,dufay,xx} and steps create  bunches. We studied set of equations (\ref{vn}) repeated 
periodically, starting from configuration of steps slightly moved in random from their initially regular positions similarly as in Ref \cite{xx}. On solving equations  (\ref{vn}) we approximated average velocity locally on setting $v_0=0.5(v_{n-1}+v_n)$. We checked that leaving $v_0$ as global mean velocity value does not changes main results much. 
Depending on the  parameters $\rho$ and $s$ steps either even out relative distances  during their evolution given by (\ref{vn}) or for other parameters create  pair or bunches. When the step to step distance becomes zero we set for them the same, average velocity. It is the same as long as according to (\ref{vn}) their distance starts to increase.  

Let us  first analyze what happens with the system with the  slow surface saturation i.e. small $s$ value. When  $s<1$  mean diffusion path of the particle over the terrace is longer than the mean inter--step distance. In such case  particles reach next step and can attach to is before it evaporates. For wider steps more particles have chance to desorb, for narrower terraces less. When the directions up and down  are not   equivalent due to the step movement this  results in the further  growth of  wider terrace and diminishing of the narrower.  
In Fig. 7 evolution  of  50 interacting steps with periodic boundary conditions was plotted for $s^2=0.09$ the same as for simulated system in Fig. 1. 
At the beginning steps were located regularly with distances $L_0$.  We  shift them randomly to the left or to the right by modifying their positions within range $\pm 0.1 L_0$. Starting from this  perturbed arrangement steps move backward approaching closer one to another. Note that in order to compare with simulation results we plot them in such a way that on coming up position in this plot decreases, hence steps with negative velocity  move upward.  

During the evolution that can be seen  in Fig. 7 at the beginning some of steps slow down some and other go faster. As an effect at around $500$ time units steps start stick together and  subsequent steps  attach to them later. Step bunches start form at around $t=600t_0$. Further evolution is realized by single steps detaching  from the  bunches and attaching  to the next neighboring ones. The evolution has similar character to this observed during Monte Carlo simulation  illustrated in Fig. 3.    
We started with the system of slightly perturbed step distances then the evolution was determined by Eqs (\ref{vn}). It can steps had tendency to stick together.  
To check whether such behavior is caused by the term responsible for the step flow the same equations  were solved on setting velocity term $v_0=0 $. Dashed lines  in Fig. 7 present results of calculation. It can be seen that  at the beginning both evolutions with the velocity term and without it are similar, but then the first leads to sticking steps together whereas according to the second step distances start to even out. Eventually steps are equidistant and move with the same velocities. Note that the mean value of step velocity is the same in both cases.  This is because the parameter $v_0$ has positive and negative incomes at right side of Eqs (\ref{vn}). As an effect it changes step velocity only locally. 
 We assumed zero Schwoebel barrier and no external particle drift. It can be seen that it is the step flow which induces the step bunching phenomenon. When we switched it off no bunching was observed. 

We can now see what happens when the sublimation is quicker i.e. at higher temperatures. Figure 8 shows step evolution for $s^2=1.9$. We can see that steps still stick together. At the first stage of evolution some steps come closer, then separate. The time evolution pattern is not so well ordered. At this part it resembles a little the one in Fig. 4, where fast sublimation process was simulated.  After that more ordered part of evolution happens with predominant step pairs. Double step structures evolve by exchanging singe steps between them. When $s$ is increased further above $s^2=2.5$, steps do not stick at all. Their evolution looks similar to the one with $v_0=0$ as plotted by dashed lines in Fig 7.

\section{Summary}
\label{sec:D}
We studied evolution of patterns at stepped surface of sublimated crystal. Simulated lattice gas model of GaN(0001) was defined by  parameters describing inter--particle interactions,  diffusion and desorption probabilities. For wide range of parameters step create bunches even if we assume that step edge-barrier is absent. At low desorption probabilities bunches include several steps and are well separated. They move slowly and are rather stiff. Single steps detach from bunches and attach to the next one moving much faster than many steps together. 
As a result surface profile has characteristic  convex form.
 We wrote BCF equations for 1D annealed system and showed that according to them for some parameters steps have tendency to bunch together when mean step velocity is taken into account. This effect vanishes when mean velocity is set to zero. We conclude that step bunching in studied systems is induced by the step flow. When low Schwoebel barrier positive or negative is present the system behavior does not change much. The only difference is that process can happen more or less slowly. 
However when barrier becomes comparable to the particle-particle  interaction constant steps bend and the system builds up as a rough surface. Such behavior of simulated surfaces happens in  two dimensions and as such cannot be described by 1D models.  It also suggests that freedom of particles to wander over the surface and jump down and up may be necessary condition for  steps to flow steadily and keep their stability.

\section{Acknowledgement}
 Research supported by the National Science Centre(NCN) of Poland (Grant NCN No. 2011/01/B/ST3/00526)

\end{document}